\documentclass[
prd 
,showpacs ,amssymb,superscriptaddress,aps
]{revtex4}
\usepackage{graphicx}
\input{epsf}

\usepackage{amsmath,amssymb}
\usepackage{bm}
\usepackage{times}

\newcommand{\dalm}{\kern1pt\vbox{\hrule height 0.9pt\hbox{\vrule width 0.9pt
\hskip 2.5pt\vbox{\vskip 5.5pt}\hskip 3pt\vrule width 0.3pt}\hrule height 0.3pt}
\kern1pt}

\newcommand{\be}{\begin{eqnarray}}
\newcommand{\ee}{\end{eqnarray}}
\newcommand{\beq}{\begin{eqnarray}}
\newcommand{\eeq}{\end{eqnarray}}
\newcommand{\pd}{\partial}

\newcommand{\nn}{\nonumber}

\begin{document}



\title{Properties of an electrically charged black hole in Eddington-inspired Born-Infeld gravity}

\author{Hajime Sotani}
\email{sotani@yukawa.kyoto-u.ac.jp}
\affiliation{Division of Theoretical Astronomy, National Astronomical Observatory of Japan, 2-21-1 Osawa, Mitaka, Tokyo 181-8588, Japan}

\author{Umpei Miyamoto}
\affiliation{Research and Education Center for Comprehensive Science, Akita Prefectural University, Akita 015-0055, Japan}

\date{\today}

\begin{abstract}
We systematically examine the properties of an electrically charged black hole in Eddington-inspired Born-Infeld gravity with not only the positive but also the negative coupling constant in the theory. As a result, we numerically find that the black hole solution exists even with the negative coupling constant, where the electric charge of black hole can be larger than the black hole mass. We also clarify the parameter space where the black hole solution exists. On the other hand, to examine the particle motion around such black hole, we derive the geodesic equation. The behavior of the effective potential for the radial particle motion is almost the same as that in general relativity, but the radius of the innermost stable circular orbit and the angular momentum giving the innermost stable circular orbit can be changed, depending on the coupling constant. In particular, we find that the radius of innermost stable circular orbit with the specific value of the coupling constant can be smaller than that for the extreme case in general relativity. Such a particle can release the gravitational binding energy more than the prediction in general relativity, which could be important from the observational point of view. 
\end{abstract}

%
\maketitle
\section{Introduction}
\label{sec:I}

Many experiments have been done to verify the gravitational theory, and all of them did not show the defectiveness of general relativity proposed by Einstein.  However, such experiments are done in a weal-field regime such as the Solar System \cite{W1993}, while the tests of gravitational theory in a strong-field regime are still quite poor. That is, the gravitational theory describing the phenomena in a strong-field regime might be different from general relativity. If so, there might be an imprint of the modified gravity in and around compact objects, i.e., one could probe the gravitational theory in a strong-field regime via astronomical observations associated with compact objects. In fact, the technology is developing more and more, which enables us to observe the compact objects and the phenomena around them with high accuracy. These observations will help us to reveal the gravitational theory in a strong-field regime \cite{P2008,SK2004,S2009,YYT2012}.

Another reason why the alternative gravitational theories are considered is the possibility to solve the observational and/or theoretical problems with general relativity. In fact, Eddington-inspired Born-Infeld gravity (EiBI) has recently attracted attention in the context to avoid the big bang singularity \cite{AF2012,BFS12013} and the singularity during the gravitational collapse of dust \cite{PCD2011}. EiBI proposed by Ba\~nados and Ferreira \cite{Banados:2010ix} is based on the Eddington action \cite{E1924} and on the nonlinear electrodynamics of Born and Infeld \cite{BI}, which becomes completely equivalent to general relativity in vacuum. EiBI is treated exactly following a Palatini approach, i.e., the metric and the connection are considered as independent fields. Since EiBI can deviate from general relativity only with the non-zero energy-momentum tensor, one can expect the significant deviation in the compact objects. Up to now, several attempts to examine the structures of compact objects in EiBI are done \cite{PCD2011,PDC2012,SLL2013,HLMS2013,Sotani:2014goa}, which shows the deviations in stellar properties from that predicted with general relativity. In addition, the stellar oscillations in relativistic stars in EiBI are also examined \cite{SLL2012,Sotani:2014xoa}, where the stability of compact stars in EiBI and the prospect to distinguish EiBI from general relativity are discussed.

Compared to the researches about the compact stars in EiBI, the studies of the black hole in EiBI are very poor. The simplest black hole solution must be the spherically symmetric one in EiBI in vacuum, which agrees with the Schwarzschild black hole in general relativity, due to the nature of EiBI. On the other hand, Ba\~nados and Ferreira have derived the electrically charged black hole solution in EiBI in their original paper of EiBI, assuming the asymptotically flatness \cite{Banados:2010ix}, which corresponds to the Reissner-Nordstrom solution in general relativity. Their solution is given by the integral form, where they only pointed out that the electrical field becomes everywhere regular with the positive coupling constant in EiBI. Recently, the possibility to exist the geonic black holes in EiBI is also suggested, where the central singularity is replaced with a wormhole supported by the electric field \cite{ORS2014}. Additionally, during we prepare this manuscript, the strong gravitational lensing with the electrically charged black hole in EiBI is discussed \cite{Wei:2014dka}, where they consider only the cases with the positive coupling constant in EiBI. Anyway, examinations about even the electrically charged black hole in EiBI are still insufficient. Thus, in this paper, we systematically examine the properties of electrically charged black hole in EiBI with the both positive and negative coupling constant in EiBI, where especially focus on the parameter space so that the black hole solution can exist. We also probe the innermost stable circular orbit around the black hole by deriving the geodesic equation for a mas sive test particle, which is one of the important properties describing the black hole. In this paper, we adopt geometric units, $c=G=1$, where $c$ and $G$ denote the speed of light and the gravitational constant, respectively, and the metric signature is $(-,+,+,+)$.

\section{Electrically Charged Black Hole in EiBI}
\label{sec:II}

The action describing EiBI is given by
\begin{equation}
  S = \frac{1}{8\pi\kappa}\int d^4x \left(\sqrt{|g_{\mu\nu} + \kappa R_{\mu\nu}|} - \lambda\sqrt{-g}\right) + S_{\rm M}[g,\Psi_{\rm M}], \label{eq:action}
\end{equation}
where $\kappa$ is the Eddington parameter with the dimensions of length squared and $\lambda$ is a dimensionless constant associated with the cosmological constant $\Lambda$ as $\lambda = 1+ \kappa\Lambda$ \cite{Banados:2010ix}. $R_{\mu\nu}$ is the symmetric part of the Ricci tensor constructed with the connection $\Gamma^\mu_{\alpha\beta}$, $S_{\rm M}$ denotes the action of matter fields coupling only to the metric, and $|g_{\mu\nu} + \kappa R_{\mu\nu}|$ means the absolute value of the determinant of the matrix of $(g_{\mu\nu} + \kappa R_{\mu\nu})$. 
Additionally, that in the limit of $\kappa=0$ also reduces to the Einstein-Hilbert action \cite{Pani:2012qd,Harko:2014oua}. That is, EiBI in the limit of $S_{\rm M}=0$ and/or $\kappa=0$ coincides with general relativity with a cosmological constant. In EiBI, the connection $\Gamma^\mu_{\alpha\beta}$ is considered as the independent field of the metric $g_{\mu\nu}$, in a similar way to the Palatini formalism of general relativity. Regarding the Eddington parameter $\kappa$, there are several constraints from the solar observations, big bang nucleosynthesis, and the existence of neutron stars \cite{Banados:2010ix,Pani:2011mg,Casanellas:2011kf,Avelino:2012ge}. Furthermore, the prospects to observationally constrain $\kappa$ are also suggested with the simultaneous measurements of the stellar radius of the neutron star with $0.5M_\odot$ and the neutron skin thickness \cite{Sotani:2014xoa} and with the observations of neutron star oscillations \cite{Sotani:2014goa}.

The field equations can be obtained by varying the action (\ref{eq:action}) with respect to the metric $g_{\mu\nu}$ and the connection $\Gamma^\mu_{\alpha\beta}$ \cite{Banados:2010ix};
\begin{gather}
  \Gamma^{\mu}_{\alpha\beta} = \frac{1}{2}q^{\mu\nu}\left(q_{\nu\alpha,\beta} + q_{\nu\beta,\alpha} - q_{\alpha\beta,\nu}\right), \\
  q_{\mu\nu} = g_{\mu\nu} + \kappa R_{\mu\nu}, \label{eq:qg} \\
  \sqrt{-q}q^{\mu\nu} = \lambda\sqrt{-g}g^{\mu\nu} - 8\pi\kappa\sqrt{-g}T^{\mu\nu}, \label{eq:qgT} 
\end{gather}
where $q_{\mu\nu}$ is an auxiliary metric and $T^{\mu\nu}$ is the standard energy-momentum tensor. We remark that $q^{\mu\nu}$ is the matrix inverse of $q_{\mu\nu}$, which is different from $g^{\mu\alpha}g^{\nu\beta}q_{\alpha\beta}$ if $T^{\mu\nu}\neq 0$, while raising and lowering indices in $T_{\mu\nu}$ should be done with the physical metric $g_{\mu\nu}$.
It is known that the auxiliary metric $q_{\mu\nu}$ dynamically approaches physical metric $g_{\mu\nu}$ in the case of $S_{\rm M}=0$ \cite{Banados:2010ix,DS2012}.

Now, we consider the spherically symmetric electrically charged black hole solution. Since such a solution has already been derived in Refs. \cite{Banados:2010ix,Wei:2014dka}, here we briefly mention it. The metric describing the spherically symmetric objects can be written as
\begin{gather}
  g_{\mu\nu}dx^\mu dx^\nu = -\psi^2 f dt^2 + f^{-1} dr^2 + r^2 d\Omega^2, \label{eq:metric1} \\
  q_{\mu\nu}dx^\mu dx^\nu = -G^2 F dt^2 + F^{-1} dr^2 + H^2 d\Omega^2, \label{eq:metric2}
\end{gather}
where $\psi$, $f$, $G$, $F$, and $H$ are functions of $r$, while $d\Omega^2=d\theta^2 + \sin^2 \theta d\phi^2$. Thus, the four velocity of a static observer is given as $u^{\mu}=\left((\psi\sqrt{f})^{-1},0,0,0\right)$. As an electromagnetic field, we simply add ${\cal L_{\rm M}}=-F_{\mu\nu}F^{\mu\nu}/16\pi$, which leads to
\begin{equation}
  T_{\mu\nu} = \frac{1}{4\pi}\left[g^{\alpha\beta}F_{\mu\alpha}F_{\nu\beta} - \frac{1}{4}g_{\mu\nu}F_{\alpha\beta}F^{\alpha\beta}\right].
\end{equation}
As mentioned before, in the limit of $T_{\mu\nu}$=0, EiBI is completely equivalent to general relativity, where the physical metric reduces to the Schwarzschild-de Sitter spacetime in general relativity, i.e.,
\begin{gather}
  \psi(r) = 1, \\ 
  f(r) = 1-2M/r - \Lambda r^2/3,
\end{gather}
while, in the limit of $\kappa=0$ with $T_{\mu\nu}\neq 0$, $g_{\mu\nu}$ reduces to the Reissner-Nordstrom-de Sitter spacetime in general relativity, i.e.,
\begin{gather}
  \psi(r) = 1, \\ 
  f(r) = 1-2M/r + Q^2/r^2 - \Lambda r^2/3, \label{eq:RN}
\end{gather}
where $M$ and $Q$ denote the mass and electric charge of the black hole.

As a source term, we only consider the electrical field given by the vector potential $A^{\mu}=(\Phi(r),0,0,0)$, where the Faraday tensor $F_{\mu\nu}$ is associated with the vector potential as $F_{\mu\nu}=A_{\nu,\mu}-A_{\mu,\nu}$. From the Maxwell equations without electromagnetic sources, i.e., $F^{\mu\nu}_{\ \ ;\nu}=0$, one can obtain the relation for the potential as
\begin{equation}
  \Phi' = \frac{c_0}{r^2}\psi,  \label{eq:Phi}
\end{equation}
where the prime denotes the derivative with respect to $r$, and $c_0$ is an integration constant. We remark that the electric field $E_{\mu}$ is defined as $E_{\mu}=F_{\mu\nu}u^{\nu}$, which can be expressed as
\begin{equation}
  E_\mu = \left(0,\frac{c_0}{r^2\sqrt{f}},0,0\right).
\end{equation}
Therefore, the strength of electric field, $E$, is given by $E=\left(E_\mu E^\mu\right)^{1/2}=c_0/r^2$. Since the asymptotic behavior of $E$ should be $Q/r^2$, the value of $c_0$ becomes equivalent to $Q$.

Using Eqs. (\ref{eq:qgT}), (\ref{eq:Phi}), and the relation of $c_0=Q$, one can obtain the metric function in Eq. (\ref{eq:metric2}) as
\begin{gather}
  F = f\left(\lambda - \frac{\kappa Q^2}{r^4}\right)^{-1}, \label{eq:Fr} \\
  G = \psi\left(\lambda - \frac{\kappa Q^2}{r^4}\right), \label{eq:Gr} \\
  H = r\sqrt{\lambda + \frac{\kappa Q^2}{r^4}}. \label{eq:Hr}
\end{gather}
On the other hand, from Eq. (\ref{eq:qg}), one can obtain the equations
\begin{gather}
  \frac{4G'H'}{GH} + \frac{2F'H'}{FH} + \frac{3F'G'}{FG} + \frac{2G''}{G} + \frac{F''}{F} = \frac{2}{\kappa F}\left(\frac{F}{f} -1\right), \label{eq:tt} \\
  \frac{4H''}{H} + \frac{2F'H'}{FH} + \frac{3F'G'}{FG} + \frac{2G''}{G} + \frac{F''}{F} = \frac{2}{\kappa F}\left(\frac{F}{f} -1\right), \label{eq:rr} \\
  -\frac{1}{H^2F} + \frac{F'H'}{FH} + \frac{G'H'}{GH} + \frac{H'^2}{H^2} + \frac{H''}{H} = \frac{1}{\kappa F}\left(\frac{r^2}{H^2}-1\right), \label{eq:thetatheta}
\end{gather}
where we adopt the relation of $FG=f\psi$ which is obtained from Eqs. (\ref{eq:Fr}) and (\ref{eq:Gr}). From Eqs. (\ref{eq:tt}) and (\ref{eq:rr}), one can get the relation, such as $G=c_1H'$, where $c_1$ is an integration constant.  Substituting Eqs. (\ref{eq:Gr}) and (\ref{eq:Hr}) into this relation, one obtains the metric function $\psi$, i.e.,
\begin{equation}
  \psi = \frac{c_1 r^2}{\sqrt{\lambda r^4 + \kappa Q^2}}.
\end{equation}
As mentioned before, since $\psi$ approaches 1 in the limit of $\kappa=0$ or $Q=0$, $c_1$ should be equivalent to $\sqrt{\lambda}$. Furthermore, Eq. (\ref{eq:thetatheta}) with the relation of $G=c_1H'$ can be transformed into
\begin{equation}
  \left(FHH'^2\right)' = H' + \frac{H'}{\kappa}(r^2-H^2),
\end{equation}
which leads to
\begin{equation}
  f = -\frac{r\sqrt{\lambda r^4 + \kappa Q^2}}{\lambda r^4 - \kappa Q^2} \left[\int \frac{(\Lambda r^4 - r^2 + Q^2)(\lambda r^4 - \kappa Q^2)}{r^4\sqrt{\lambda r^4 + \kappa Q^2}}dr + c_2 \right],
\end{equation}
where $c_2$ is an integration constant. With this expression, the metric function $f$ in the limit of $\kappa=0$ becomes $f = 1 - c_2/(\sqrt{\lambda}r) + Q^2/r^2 - \Lambda r^2/3$, which should be equivalent to Eq. (\ref{eq:RN}). Thus, we can fix $c_2$ to be $2\sqrt{\lambda}M$.

At last, we got the electrically charged black hole solution in EiBI as
\begin{gather}
  f = -\frac{r\sqrt{\lambda r^4 + \kappa Q^2}}{\lambda r^4 - \kappa Q^2} \left[\int \frac{(\Lambda r^4 - r^2 + Q^2)(\lambda r^4 - \kappa Q^2)}{r^4\sqrt{\lambda r^4 + \kappa Q^2}}dr + 2\sqrt{\lambda}M \right], \label{eq:fr1} \\
  \psi = \frac{\sqrt{\lambda} r^2}{\sqrt{\lambda r^4 + \kappa Q^2}}, \\
    E_\mu = \left(0,\frac{Q}{r^2\sqrt{f}},0,0\right),
\end{gather}
while the auxiliary metric functions in $q_{\mu\nu}$ are given by Eqs. (\ref{eq:Fr}) -- (\ref{eq:Hr}). We remark that the black hole solution obtained here is a generalization of the solution given by Ba\~nados and Ferreira \cite{Banados:2010ix}, corresponding to the vanishing cosmological constant $\lambda=1$ ($\Lambda=0$). In addition, as already pointed out by Ba\~nados and Ferreira \cite{Banados:2010ix}, the generalized solution (\ref{eq:fr1}) can also be expressed with an elliptic integral of first kind, such as
\begin{equation}
   f = -\frac{r\sqrt{\lambda r^4 + \kappa Q^2}}{\lambda r^4 - \kappa Q^2} \left[
      \frac{\sqrt{\lambda r^4+\kappa Q^2}(\Lambda r^4 -3r^2 + Q^2)}{3r^3} 
      + \frac{4iQ^2(\kappa\Lambda-\lambda)
      {\cal F}\left[i{\rm arcsinh}\left(r\sqrt{\frac{i\sqrt{\lambda}}{Q\sqrt{\kappa}}}\right),-1\right]}{3\sqrt{iQ\sqrt{\lambda\kappa}}}
      + 2\sqrt{\lambda}M\right],
\end{equation}
where ${\cal F[*,*]}$ denotes the elliptic integral.
Hereafter, we will simply consider only the asymptotically flat solution, i.e., $\lambda=1$ ($\Lambda=0$).

\section{Parameter Space to exist an electrically charged black hole solution in EiBI}
\label{sec:III}

\begin{figure*}
\begin{center}
\begin{tabular}{cc}
\includegraphics[scale=0.5]{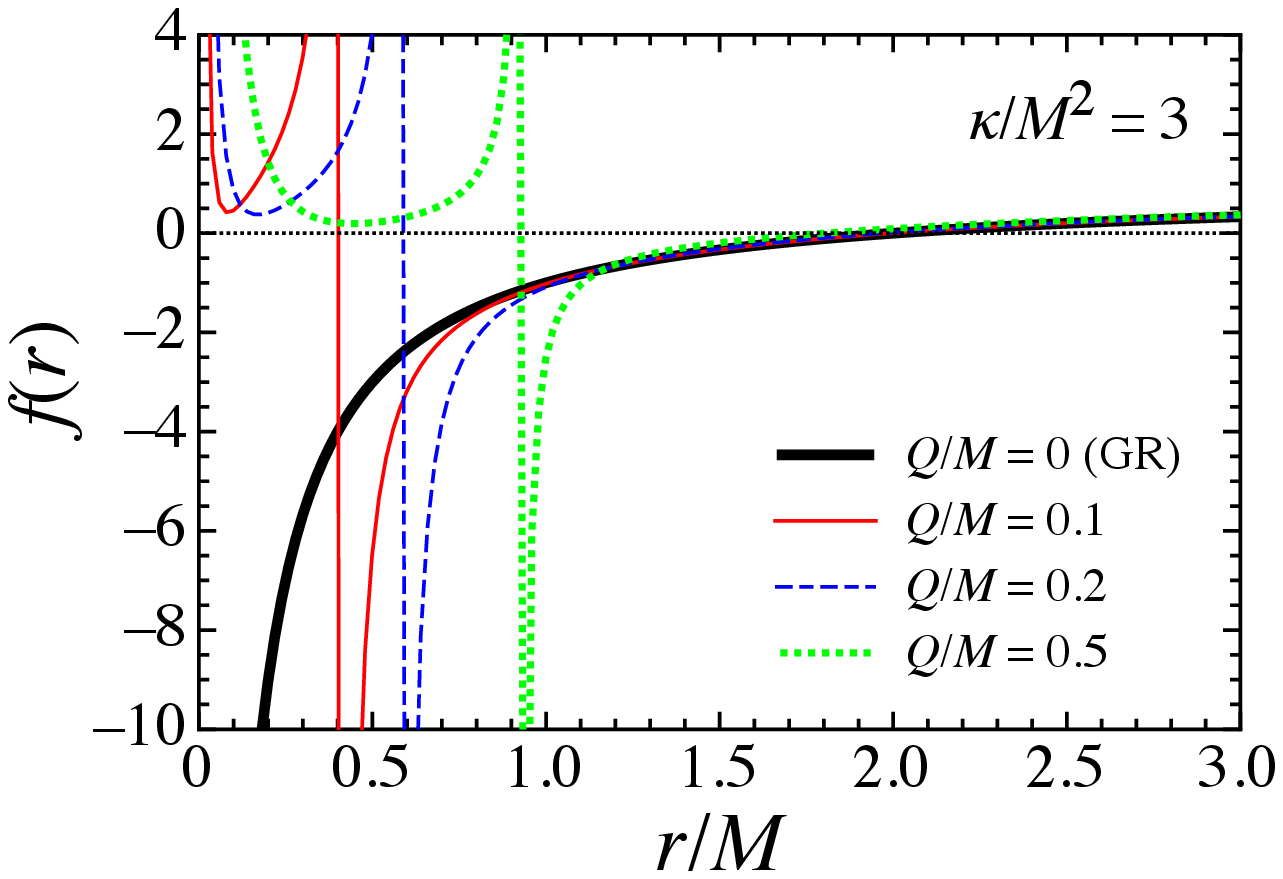} &
\includegraphics[scale=0.5]{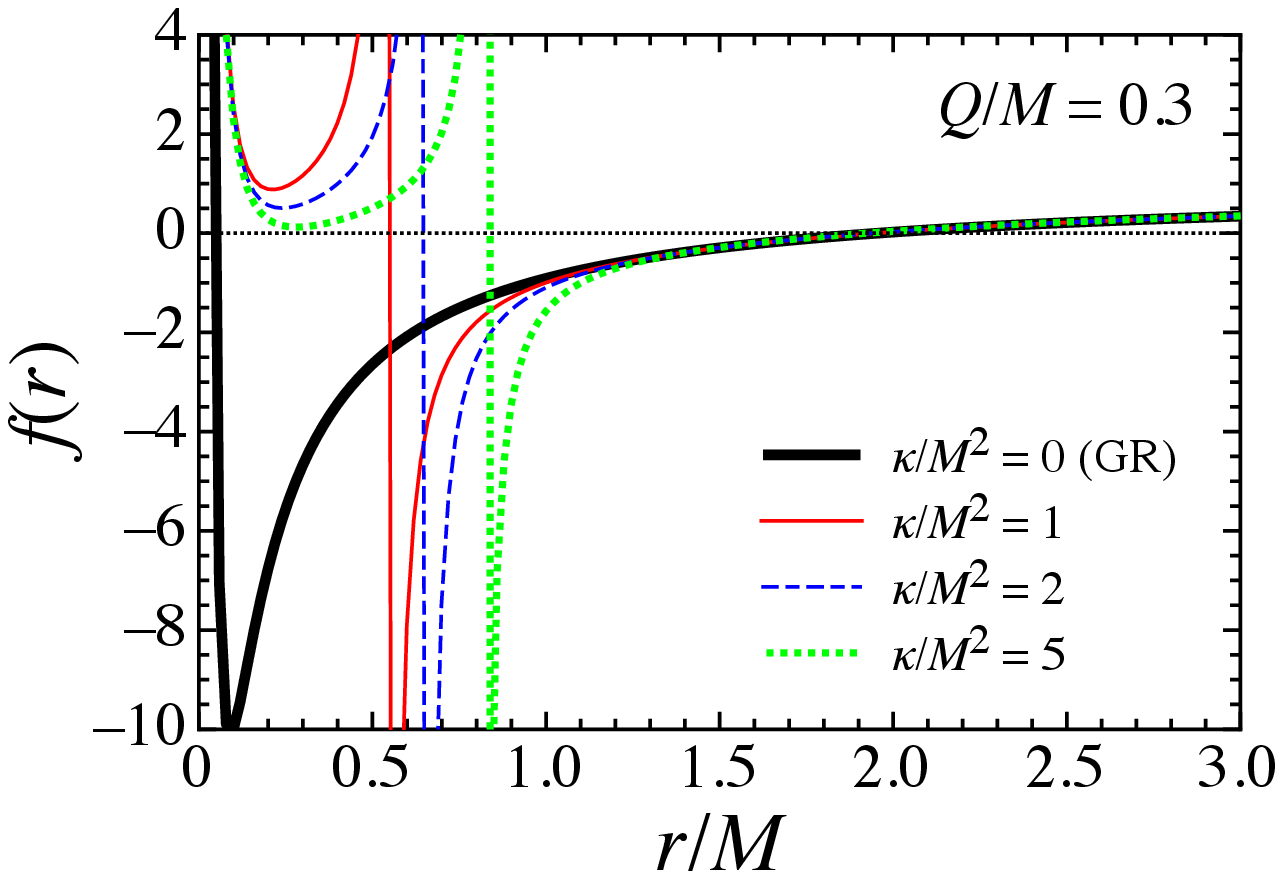}
\end{tabular}
\end{center}
\caption{
Distribution of the metric function $f(r)$ with positive $\kappa$. The left panel is the results for $Q/M=0$, 0.1, 0.2, and 0.5 with $\kappa/M^2=3$, while the right panel is those for $\kappa/M^2=0$, 1, 2, and 5 with $Q/M=0.3$. The event horizon corresponds to the rightmost position where $f(r)=0$.
}
\label{fig:fr}
\end{figure*}

As mentioned before, since the dimension of the coupling parameter $\kappa$ is length squared, $\kappa/M^2$ becomes a dimensionless parameter. The existence of neutron stars suggests the constraint on $\kappa$, such as $|\kappa|\lesssim 1$ m$^5$kg$^{-1}$s$^{-2}$ \cite{Pani:2011mg}, which leads to
\begin{equation}
  \left|\frac{\kappa}{M^2}\right| \lesssim 6.87 \times 10^3 \times \left(\frac{M_\odot}{M}\right)^2,
\end{equation}
where $M$ denotes the black hole mass considered here.
Now, to see the behavior of the metric function $f(r)$, we numerically integrate Eq. (\ref{eq:fr1}) by fixing the parameter set $(\kappa,Q)$. Figure \ref{fig:fr} shows the distribution of $f(r)$ with some parameter sets, especially for positive $\kappa$, where the left panel corresponds to the cases for $Q/M=0$, 0.1, 0.2, and 0.5 with $\kappa/M^2=3$, while the right panel corresponds to the cases for $\kappa/M^2=0,$ 1, 2, and 5 with $Q/M=0.3$. We remark that, as mentioned before, the cases of $Q=0$ and $\kappa=0$ become equivalent to the black hole solution in general relativity, i.e., the Schwarzschild solution for $(Q/M,\kappa/M^2)=(0,3)$ in the left panel and the Reissner-Nordstrom solution for $(Q/M,\kappa/M^2)=(0.3,0)$ in the right panel. From Eq. (\ref{eq:fr1}), one can find the singularity at $r=\sqrt{\sqrt{\kappa}Q}$ for positive $\kappa$, which corresponds to the vertical lines in Fig. \ref{fig:fr}. That singularity is always covered by the event horizon at least with the adopted parameter sets, where the event horizon is determined by solving the condition of $f(r)=0$. However, with larger values of $Q$ and $\kappa$, the singularity at $r=\sqrt{\sqrt{\kappa}Q}$ approaches the event horizon.
So, there must be a parameter regime where no black hole solution exists.
In fact, if one adopts extremely large values of $Q$ and $\kappa$, the singularity becomes naked.

\begin{figure*}
\begin{center}
\begin{tabular}{cc}
\includegraphics[scale=0.5]{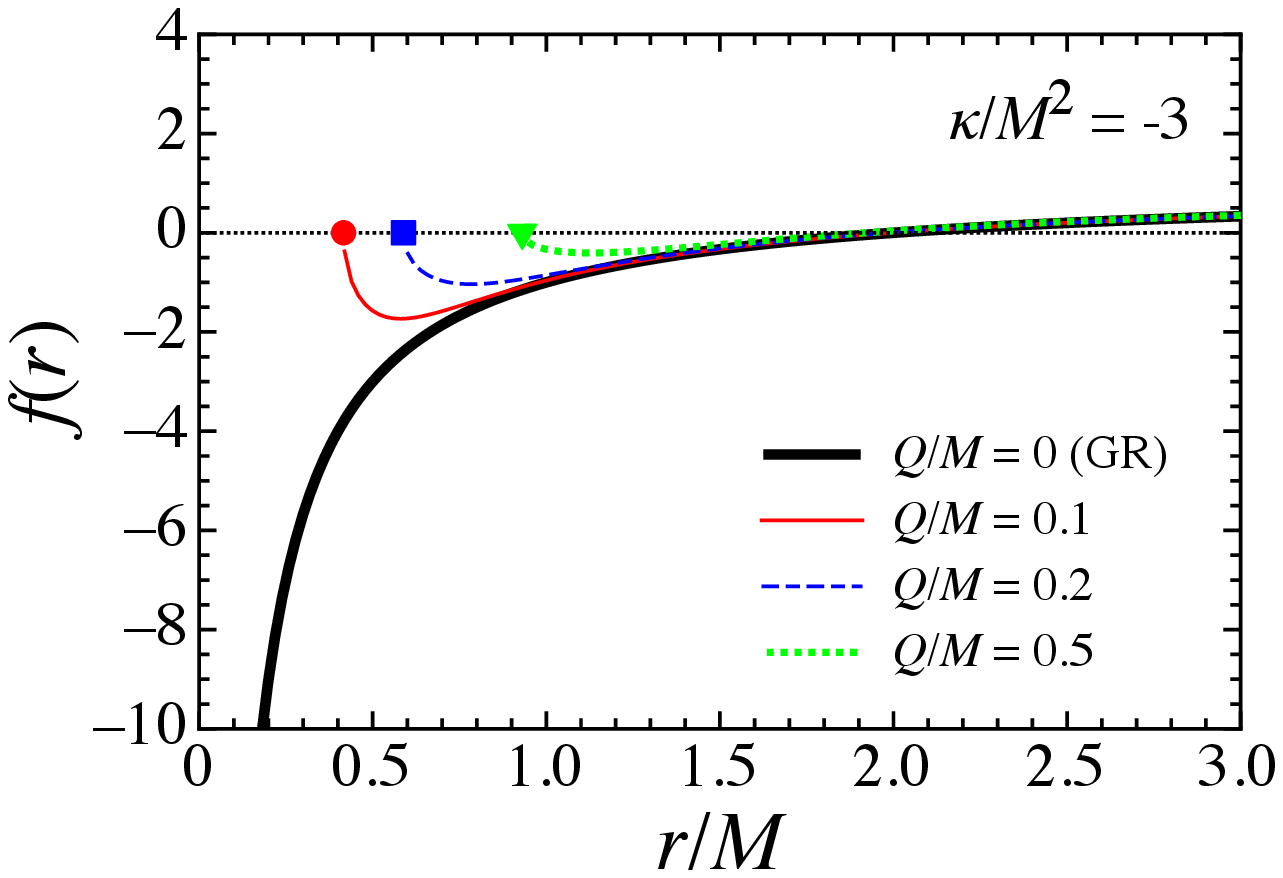} &
\includegraphics[scale=0.5]{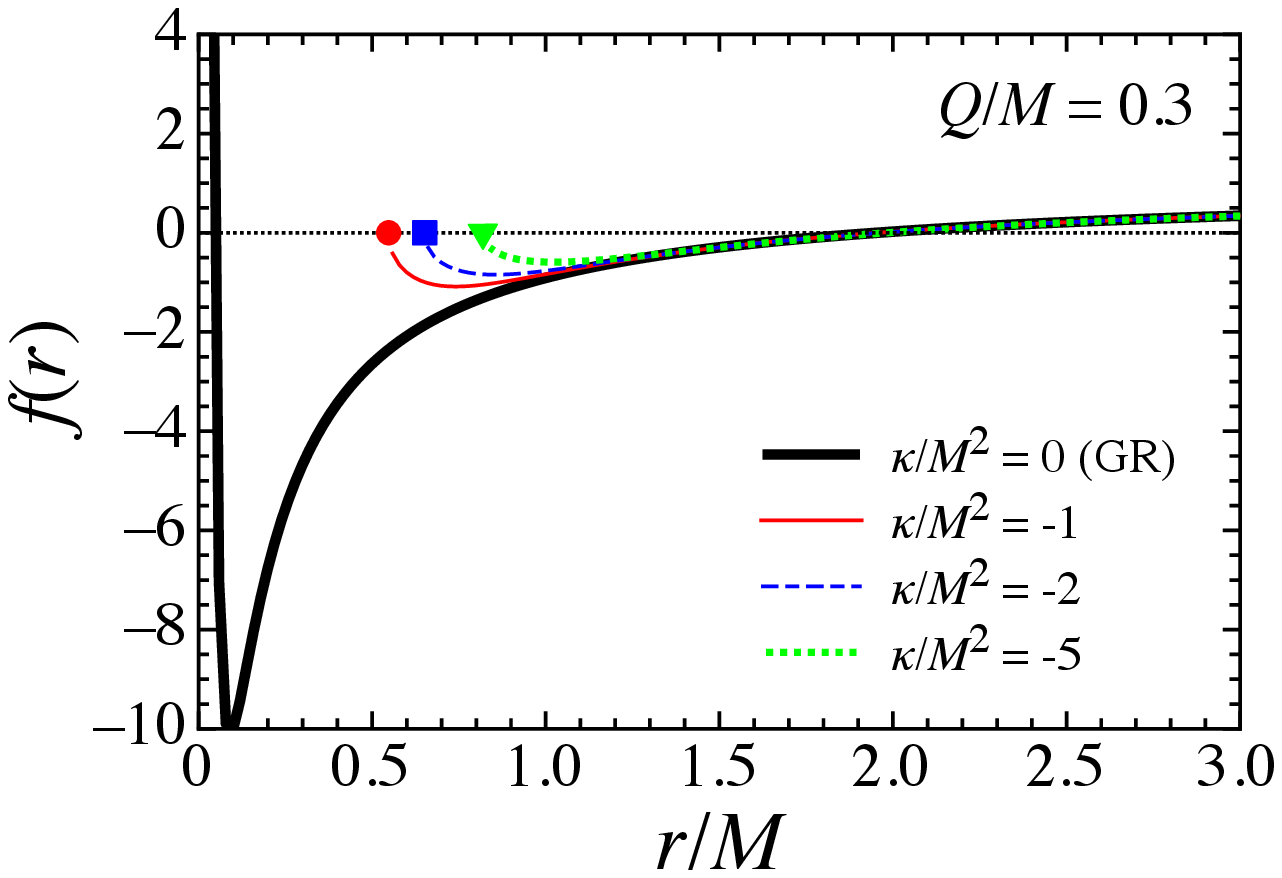}
\end{tabular}
\end{center}
\caption{
Distribution of the metric function $f(r)$ with negative $\kappa$. The left panel is the results for $Q/M=0$, 0.1, 0.2, and 0.5 with $\kappa/M^2=-3$, while the right panel is those for $\kappa/M^2=0$, $-1$, $-2$, and $-5$ with $Q/M=0.3$. The marks in the both panels denote the position of singularity for each parameter set, i.e., $r=\sqrt{\sqrt{-\kappa}Q}$.
}
\label{fig:frm}
\end{figure*}

In the similar way, Fig. \ref{fig:frm} shows the distribution of $f(r)$ for negative $\kappa$, where the left panel corresponds to the cases for $Q/M=0$, 0.1, 0.2, and 0.5 with $\kappa/M^2=-3$, while the right panel corresponds to the cases for $\kappa/M^2=0,$ $-1$, $-2$, and $-5$ with $Q/M=0.3$. From Eq. (\ref{eq:fr1}) for negative $\kappa$, one can see that the position at $r=\sqrt{\sqrt{-\kappa}Q}$ becomes singular, because the denominator in the integrand becomes zero. This position for each parameter set is shown with marks in Fig. \ref{fig:frm}. From this figure, one observes that such a singularity can be covered by the event horizon even for negative $\kappa$. Even so, as  for positive $\kappa$, since the singularity at $r=\sqrt{\sqrt{-\kappa}Q}$ approaches the event horizon with larger values of $Q$ and $|\kappa|$. 
Namely, there must be a parameter regime where no black hole solution exists. It is analytically shown that $f(r)=0$ at $r=\sqrt{\sqrt{-\kappa}Q}$ (see Appendix \ref{sec:a1} for details). 
The spacetime is singular at this surface in the sense that the scalar defined as ${\cal R}\equiv q^{\mu\nu}R_{\mu\nu}$, whose explicit form with a general value of $\lambda$ is presented below, diverges in the limit of $r \to \sqrt{\sqrt{-\kappa}Q}$,
\begin{equation}
  {\cal R}
     = \frac{4\lambda(\lambda-1)r^8 - 4\kappa^2 Q^4}{\kappa(\lambda r^4 + \kappa Q^2)(\lambda r^4 - \kappa Q^2)}.
\end{equation}

\begin{figure}
\begin{center}
\includegraphics[scale=0.53]{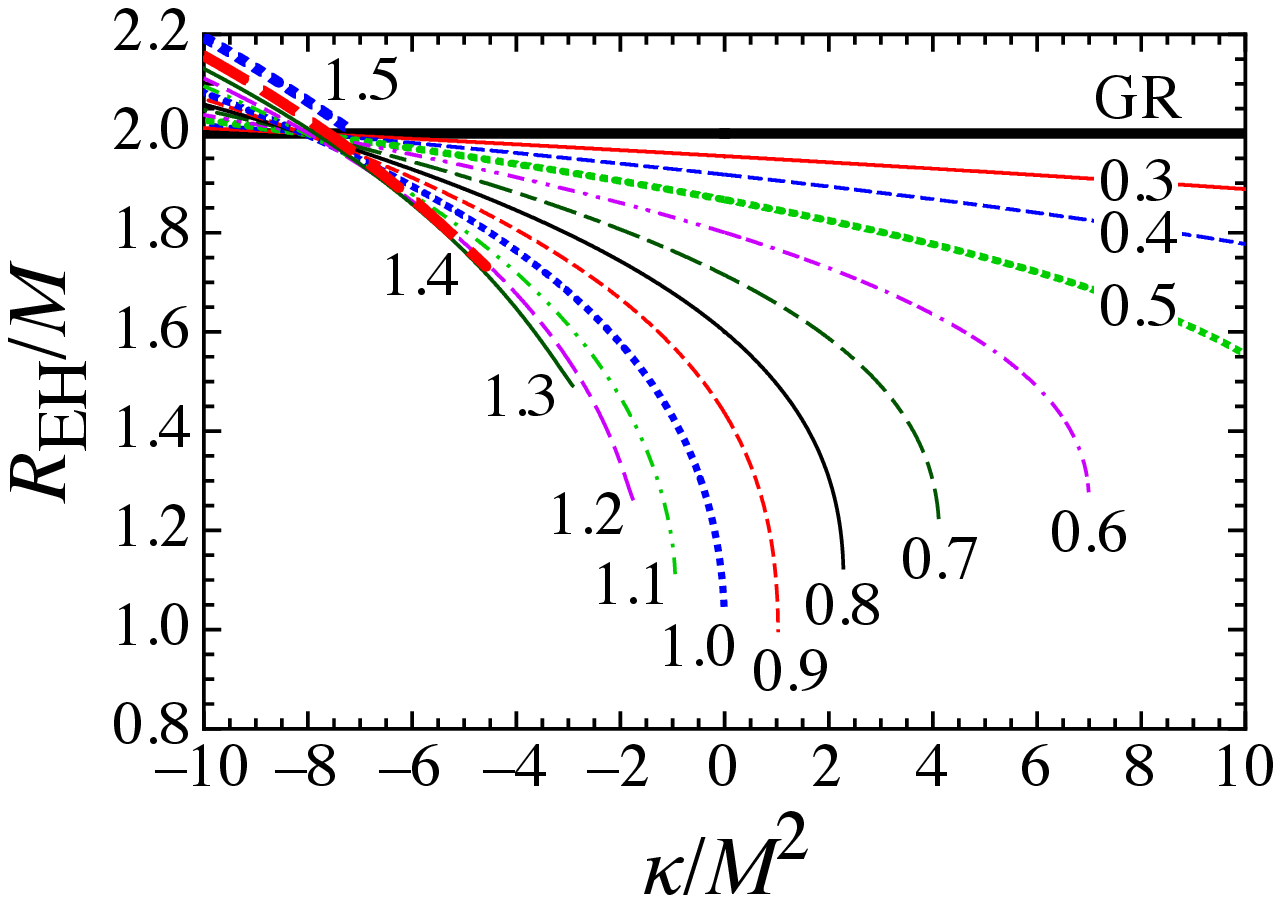} 
\end{center}
\caption{
Radius of the event horizon $R_{\rm EH}$ of an electrically charged black hole in EiBI as a function of the coupling parameter $\kappa/M^2$ with the fixed value of $Q/M$. The labels in the figure denote the fixed values of $Q/M$. In this figure, the thick-solid line corresponds to the radius of event horizon of black hole solution with $Q=0$, i.e, the Schwarzschild black hole in general relativity.
}
\label{fig:REH-kk}
\end{figure}

\begin{figure}
\begin{center}
\includegraphics[scale=0.53]{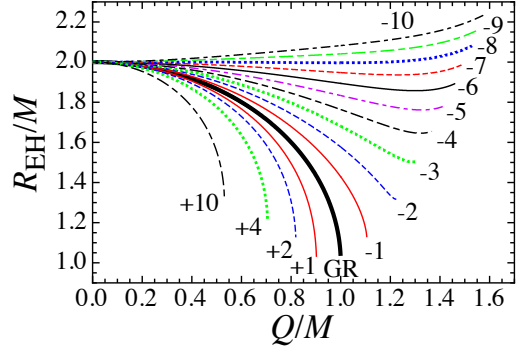} 
\end{center}
\caption{
Radius of the event horizon $R_{\rm EH}$ of an electrically charged black hole in EiBI as a function of the electrical charge $Q/M$ with the fixed value of $\kappa/M^2$. The labels in the figure denote the fixed values of $\kappa/M^2$. In this figure, the thick-solid line corresponds to the radius of event horizon of black hole solution with $\kappa=0$, i.e, the Reissner-Nordstrom black hole in general relativity.
}
\label{fig:REH-QQ}
\end{figure}

\begin{figure}
\begin{center}
\includegraphics[scale=0.53]{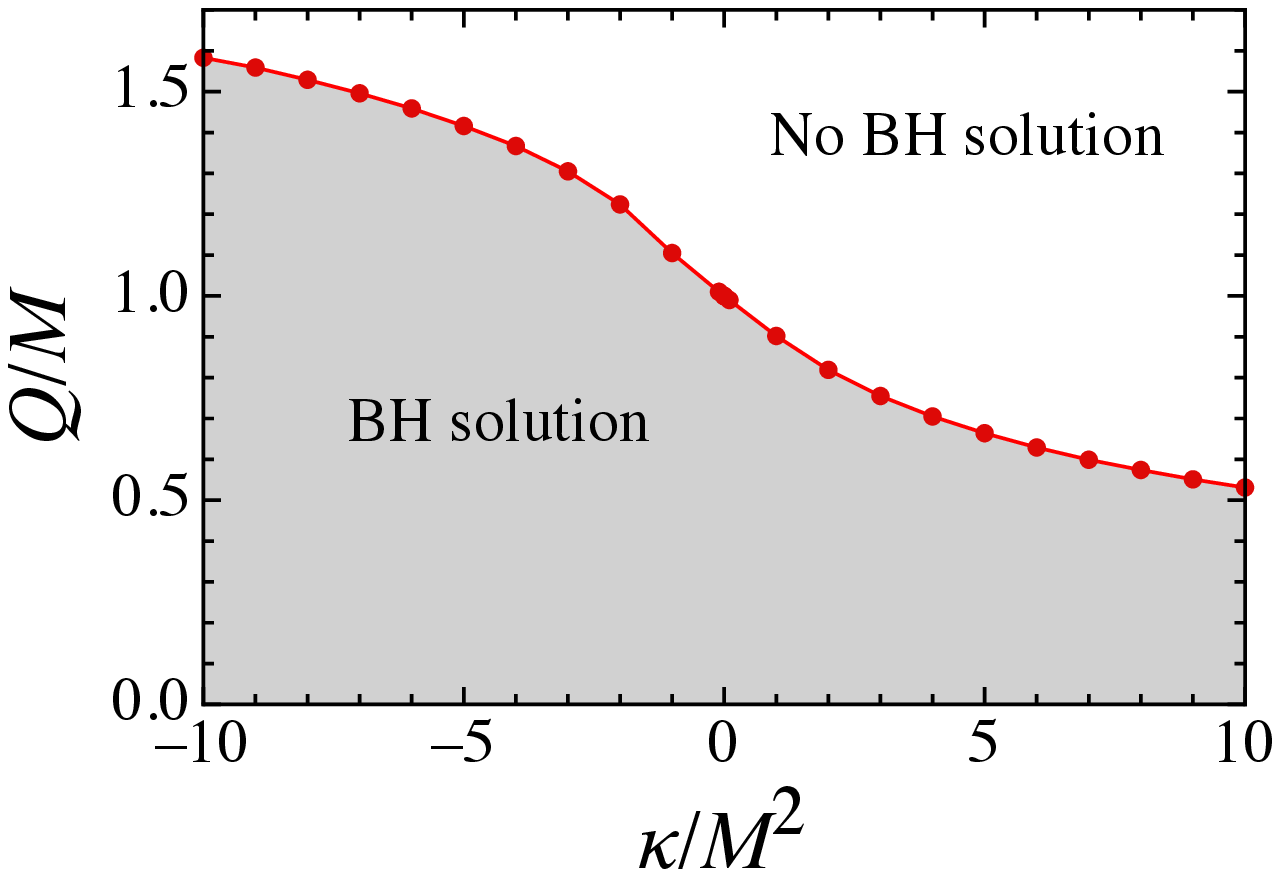} 
\end{center}
\caption{
Parameter space for an electrically charged black hole in EiBI to exist corresponds to the region below the curve (shaded region).
}
\label{fig:kk-QQ}
\end{figure}

We plot the radius of the event horizon $R_{\rm EH}/M$ in Fig. \ref{fig:REH-kk} as a function of $\kappa/M^2$ by fixing the value of $Q/M$, while in Fig. \ref{fig:REH-QQ} as a function of $Q/M$ by fixing the value of $\kappa/M^2$. We remark that the thick-solid lines in Figs. \ref{fig:REH-kk} and \ref{fig:REH-QQ} correspond to the results in general relativity, but different black hole solutions from each other, i.e., the Schwarzschild black hole (for $Q/M=0$ in EiBI) in Fig. \ref{fig:REH-kk}, while the Reissner-Nordstrom black hole (for $\kappa/M^2=0$ in EiBI) in Fig. \ref{fig:REH-QQ}. From these figures, we find that the black hole solution can exist even for $Q/M>1$ with negative $\kappa$, although $Q/M=1$ is the extreme case in general relativity. On the other hand, as the value of positive $\kappa$ increases, the maximum value of charge 
where the black hole solution exists decreases.
Additionally, from Fig. \ref{fig:REH-QQ}, we find that the radius of the event horizon in EiBI with fixed value of $Q/M$ decreases as the positive $\kappa$ increases, while increases as the negative $\kappa$ decreases, compared with the prediction in general relativity. In practice, the radii of event horizon with $Q/M=0.5$ in EiBI are $R_{\rm EH}/M = 1.85$, $1.83$, and $1.78$ for $\kappa/M^2=+1$, $+2$, and $+4$, while $R_{\rm EH}/M = 1.89$, $1.90$, and $1.94$ for  with $\kappa/M^2=-1$, $-2$, and $-4$, where the radius of event horizon with $Q/M=0.5$ in general relativity is $R_{\rm EH}/M=1.87$. We also remark from Fig. \ref{fig:REH-kk} that the radius of the event horizon with $\kappa/M^2$ smaller than around $-8$ becomes always larger than that of Schwarzschild black hole, independently of the value of charge. Finally, we show the parameter space 
where an electrically charged black hole solution exists in EiBI in Fig. \ref{fig:kk-QQ}, where the region below the curve corresponds to such a parameter space.

\section{Geodesic equation on electrically charged black hole solution in EiBI}
\label{sec:IV}
We consider the geodesic equation on the electrically charged black hole spacetime in EiBI given in \S \ref{sec:II}. 
The Lagrangian for a free particle is
\be
	{\cal L}=
	g_{\mu \nu} \frac{dx^\mu}{d\lambda} \frac{dx^\nu}{d\lambda}.
\label{Lagrangian1}
\ee
Varying the action $ \int {\cal L} d\lambda  $ with respect to $x^\mu$, we obtain the Euler-Lagrange equation,
\be
		\frac{\pd {\cal L}}{\pd x^\mu} - \frac{d}{d\lambda}\left( \frac{\pd {\cal L}}{\pd \dot{x}^\mu} \right) = 0,
\ee
where the dot denotes the derivative with respect to $\lambda$. Substituting the metric ansatz \eqref{eq:metric1} into the Lagrangian \eqref{Lagrangian1}, we obtain
\be
	{\cal L}
	=
	- \psi^2 f \dot{t}^2 + f^{-1} \dot{r}^2 + r^2 ( \dot{\theta}^2 +  \sin^2 \theta \dot{\phi}^2 ).
\ee

Since $t$ and $\phi$ are the cyclic coordinates, the $t$- and $\phi$-components of the geodesic equations can be integrated once to give
\be
	\dot{t} = \frac{e}{f \psi^2} \ \ \ {\rm and}
\ \ \ 
	\dot{\phi} = \frac{\ell}{r^2 \sin^2 \theta},
\ee
where $e$ and $\ell$ are integration constants corresponding to the energy and angular momentum (per unit rest mass in the present massive case), respectively.

We can assume that the particle motion is confined in the equatorial plane ($\theta \equiv \pi/2$) due to the nature of spherically symmetric spacetime. Moreover, since the Lagrangian is conserved along the geodesic, i.e., $d{\cal L}/d\lambda = 0$, we put ${\cal L} = -1 $ for a massive particle using the degree of freedom of the affine parameter $\lambda$ (see Appendix \ref{sec:a2} for a massless particle). Thus, we obtain a potential problem for the radial motion of particle as
\be
	\psi^2\left( \frac{dr}{d\lambda} \right)^2 + V^2 = e^2,
\;\;\;
	V(r) = \sqrt{f\psi^2\left(1+\frac{\ell^2}{r^2} \right)}.
\ee
Thus, if one considers the circular orbit of a massive test particle around an electrically charged black hole in EiBI, $dr/d\lambda$ should be zero, which leads to the condition of $e=V(r)$, and such orbit realizes at the minimum point of the effective potential $V(r)$.

\begin{figure}
\begin{center}
\includegraphics[scale=0.53]{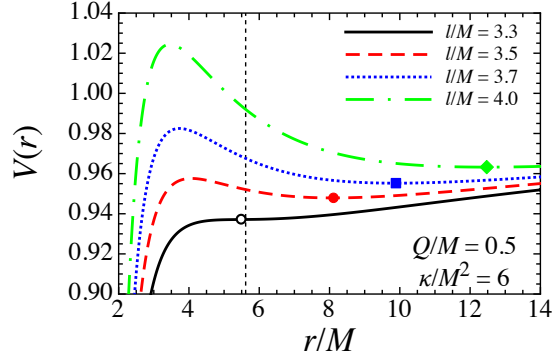} 
\end{center}
\caption{
Effective potential $V(r)$ in EiBI with the various values of $\ell/M$ for $Q/M=0.5$ and $\kappa/M^2=6$, where the solid, broken, dotted, and dot-dashed curves correspond to the cases with $\ell/M=3.3$, $3.5$, $3.7$, and $4.0$, respectively, while the mark on each curve denotes the minimum point of the effective potential. In particular, the open circle corresponds to the radius of ISCO. The vertical dashed line denotes the position of ISCO for the Reissner-Nordstrom black hole with $Q/M=0.5$ in general relativity.
}
\label{fig:Vr}
\end{figure}

As an example, we plot the distributions of $V(r)$ with various values of $\ell/M$ for $Q/M=0.5$ and $\kappa/M^2=6$ in Fig. \ref{fig:Vr}, where the solid, broken, dotted, and dot-dashed curves correspond to the cases with $\ell/M=3.3$, $3.5$, $3.7$, and $4.0$, respectively, while the mark on each curve denotes the minimum point of the effective potential for each case. Thus, these minimum points correspond to the radius of the circular orbit of the test particle. From this figure, one can observe that, as $\ell/M$ decreases, the maximum of the effective potential also decreases and the position of the minimum point shifts to left, which is similar to the general relativity case. That is, as in general relativity, there exists the minimum value of $\ell/M$, with which the circular orbit of the test particle can exist. For this minimum value of $\ell/M$, the radius of circular orbit also becomes minimum, which corresponds to the innermost stable circular orbit (ISCO). In Fig. \ref{fig:Vr}, we additionally show the position of ISCO for the Reissner-Nordstrom black hole with $Q/M=0.5$ in general relativity ($\kappa=0$) with the vertical dashed curve, while the open circle denotes the position of ISCO in EiBI with $Q/M=0.5$ and $\kappa/M^2=6$. From this picture, the position of ISCO can deviate from that expected in general relativity, depending on the coupling constant in EiBI. Furthermore, we find that the minimum value of $\ell/M$ giving ISCO also depends on the gravitational theory, i.e, $\ell_{\rm min}/M=3.30$ in EiBI with $Q/M=0.5$ and $\kappa/M^2=6$, while $\ell_{\rm min}/M=3.34$ in general relativity with $Q/M=0.5$. We remark that the radius of ISCO for the Schwarzschild black hole is $R_{\rm ISCO}/M=6$ for $\ell/M=2\sqrt{3}$.

\begin{figure}
\begin{center}
\includegraphics[scale=0.53]{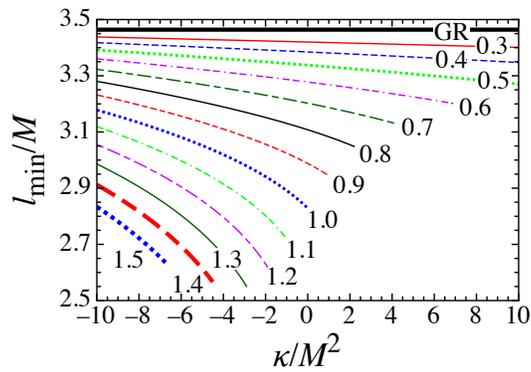} 
\end{center}
\caption{
Minimum value of $\ell/M$, which corresponds to ISCO in EiBI, as a function of the coupling parameter $\kappa/M^2$ with the fixed value of $Q/M$. The levels in the figure denote the fixed values of $Q/M$. In this figure, the thick-solid line corresponds to the radius of event horizon of black hole solution with $Q = 0$, i.e, the Schwarzschild black hole in general relativity.
}
\label{fig:Lmin-kk}
\end{figure}

\begin{figure}
\begin{center}
\includegraphics[scale=0.53]{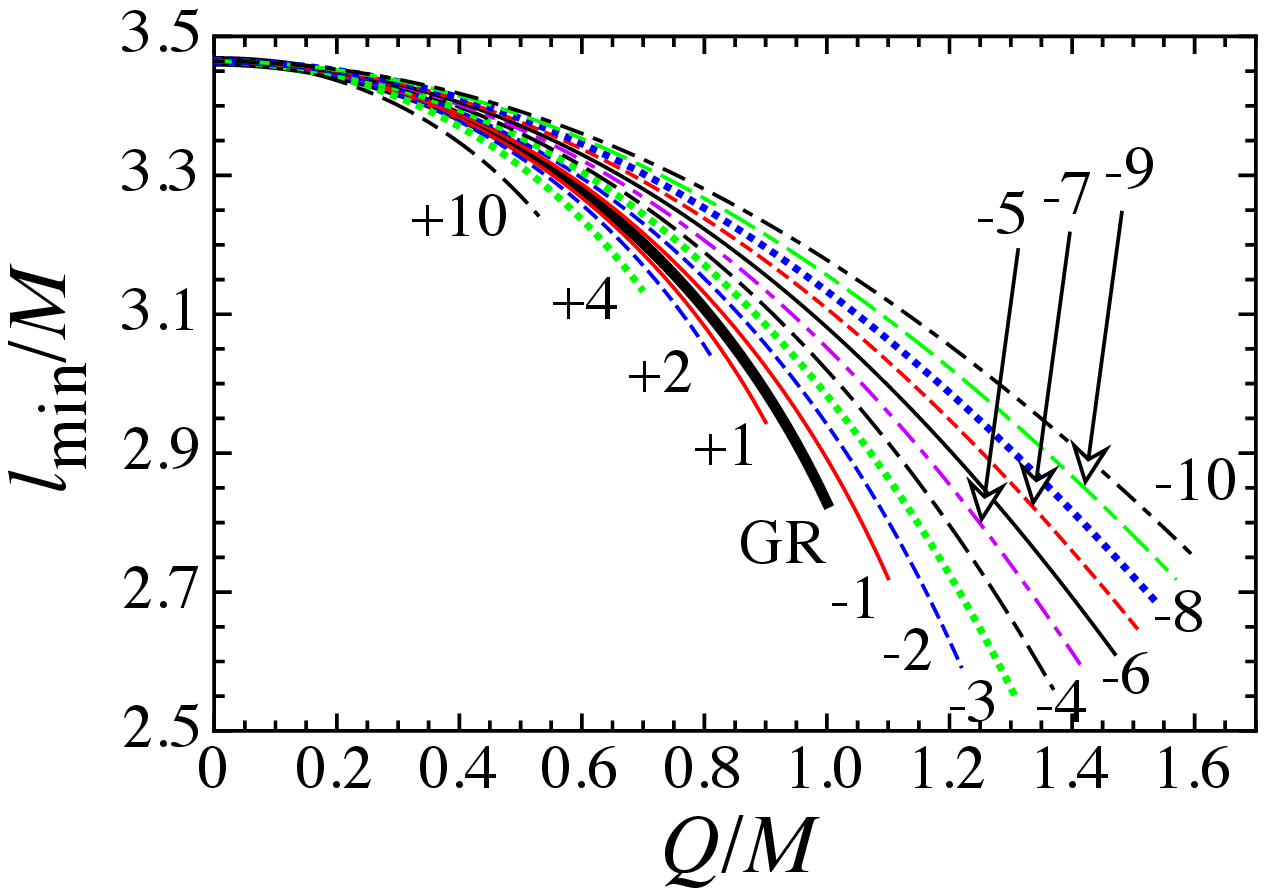} 
\end{center}
\caption{
Minimum value of $\ell/M$, which corresponds to ISCO in EiBI, as a function of the electrical charge $Q/M$ with the fixed value of $\kappa/M^2$. The labels in the figure denote the fixed values of $\kappa/M^2$. In this figure, the thick-solid line corresponds to the radius of event horizon of black hole solution with $\kappa=0$, i.e, the Reissner-Nordstrom black hole in general relativity. 
}
\label{fig:Lmin-QQ}
\end{figure}

\begin{figure}
\begin{center}
\includegraphics[scale=0.53]{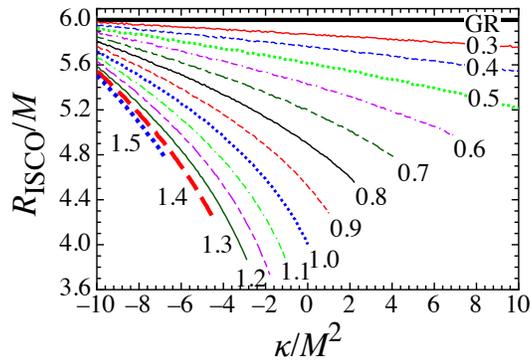} 
\end{center}
\caption{
Radius of ISCO in EiBI as a function of the coupling parameter $\kappa/M^2$ with the fixed value of $Q/M$. The levels in the figure denotes the fixed value of $Q/M$. In this figure, the thick-solid line corresponds to the radius of event horizon of black hole solution with $Q = 0$, i.e, the Schwarzschild black hole in general relativity.
}
\label{fig:Risco-kk}
\end{figure}

\begin{figure}
\begin{center}
\includegraphics[scale=0.53]{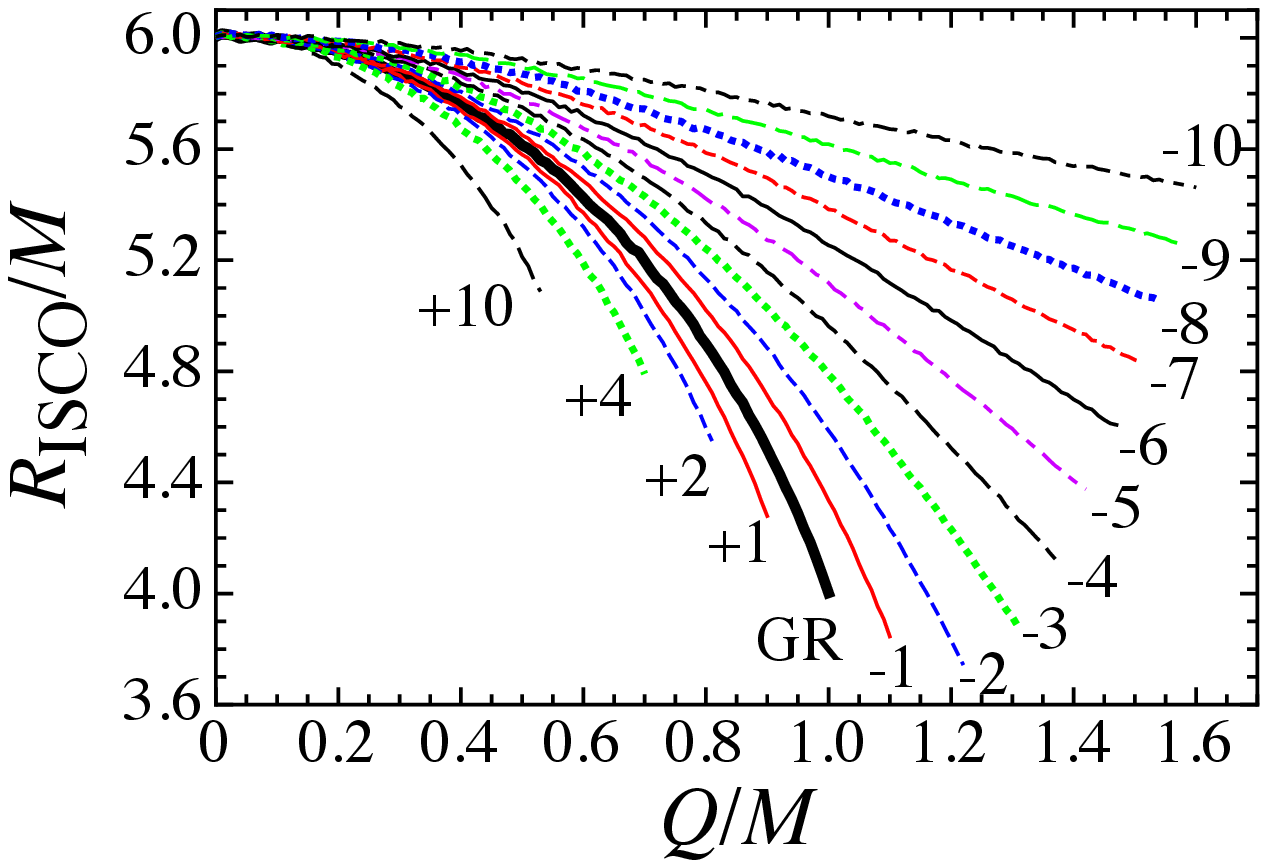} 
\end{center}
\caption{
Radius of ISCO in EiBI as a function of the electrical charge $Q/M$ with the fixed value of $\kappa/M^2$. The labels in the figure denote the fixed values of $\kappa/M^2$. In this figure, the thick-solid line corresponds to the radius of event horizon of black hole solution with $\kappa=0$, i.e, the Reissner-Nordstrom black hole in general relativity. 
}
\label{fig:Risco-QQ}
\end{figure}

In Fig. \ref{fig:Lmin-kk}, we show the minimum value of $\ell/M$ corresponding to the ISCO in EiBI with the fixed value of $Q/M$ as a function of the coupling constant $\kappa/M^2$, where the labels in the figure denote the fixed values of $Q/M$. Additionally, in Fig. \ref{fig:Lmin-QQ}, we show the similar figure to Fig. \ref{fig:Lmin-kk}, but as a function of the electrical charge $Q/M$ with the fixed value of $\kappa/M^2$, where the labels in the figure denote the fixed values of $\kappa/M^2$. From Fig. \ref{fig:Lmin-QQ}, one can see that $\ell_{\rm min}/M$ with fixed $Q/M$ decreases as the positive $\kappa/M^2$ increases, while $\ell_{\rm min}/M$ with fixed $Q/M$ increases as the negative $\kappa/M^2$ decreases, compared with the case in general relativity. Moreover, the corresponding radius of ISCO are shown in Figs. \ref{fig:Risco-kk} and \ref{fig:Risco-QQ}. Figure \ref{fig:Risco-kk} shows the relation between the radius of ISCO in EiBI with fixed value of $Q/M$ and the coupling constant $\kappa/M^2$, while Fig. \ref{fig:Risco-QQ} denotes the relation between the radius of ISCO in EiBI with fixed value of $\kappa/M^2$ and the electrical charge of black hole $Q/M$. From Fig. \ref{fig:Risco-QQ}, one can observe that the same behavior in $R_{\rm ISCO}/M$ as in Fig. \ref{fig:Lmin-QQ}, i.e., $R_{\rm ISCO}/M$ decreases as the positive $\kappa/M^2$ increases and $R_{\rm ISCO}/M$ increases as the negative $\kappa/M^2$ decreases. In practice, the radii of ISCO in EiBI with $Q/M=0.5$ are $R_{\rm ISCO}/M=5.54$, $5.47$, and $5.39$ for $\kappa/M^2=+2$, $+4$, and $+6$, while $R_{\rm ISCO}/M=5.69$, $5.75$, and $5.80$ for $\kappa/M^2=-2$, $-4$, and $-6$, where the radius of ISCO in general relativity with $Q/M=0.5$ is $R_{\rm ISCO}/M=5.61$. In addition, we find that the radius of ISCO in EiBI with $-3\lesssim \kappa/M^2<0$ and $Q/M>1$ can become smaller than that for the extreme case in general relativity. That is, the particle in EiBI can approach the central black hole closer in the specific case. If so, such a particle in EiBI has a potential to release the gravitational binding energy more than the case in general relativity.

\section{Conclusion}
\label{sec:IV}

In this article, we consider an electrically charged black hole in Eddington-inspired Born-Infeld gravity (EiBI), especially focusing on the asymptotically flat solution, and systematically examine the properties of such a black hole with not only the positive but also the negative coupling constant, $\kappa$, in EiBI. Then, we find that the black hole solution even with negative coupling constant exists. We also show numerically that the radius of event horizon in EiBI with the positive coupling constant becomes smaller than that in general relativity, while the radius of event horizon in EiBI with the negative coupling constant becomes larger than that in general relativity. In particular, it is interesting that the black hole solution with the negative coupling constant can exist even for $Q/M>1$, where $Q$ and $M$ denote the charge and mass of black hole, respectively. Additionally, we show the parameter space in $Q/M$ and $\kappa/M^2$, where the electrically charged black hole can exist. The maximum value of $Q/M$ in such a parameter space depends strongly on the coupling constant in EiBI.

We also derive the geodesic equation to examine the massive particle motion around the electrically charged black hole in EiBI. The behavior of the effective potential in the geodesic equation for the radial motion is essentially the same as that in general relativity, which suggests the existence of the innermost stable circular orbit (ISCO) in EiBI. The dependence of the radius of ISCO with the fixed coupling constant on the electrically charge is quite similar to that of the event horizon in EiBI, where we find that the radius of ISCO in EiBI can become smaller than that for the extreme case in general relativity, depending on the value of the negative coupling constant. Such a particle can release the gravitational binding energy more than the expectation in general relativity, which might be important from the observational point of view. In this article, as a first step, we focus only on the asymptotically flat solution, but we will examine the properties of black hole solution without such an assumption somewhere. Additionally, since the genuine electrically charged black hole might be difficult to find in our Universe, we should add the effect of the rotation of black hole, which will be also done somewhere.

\acknowledgments
We are grateful to the referee for reading our manuscript carefully and giving valuable comments.
This work was supported in part by Grant-in-Aid for Young Scientists (B) through No.\ 26800133 
(HS) and No.\ 25800157 (UM) provided by JSPS.

\appendix
\section{Behavior of $f(r)$ near the singularity for $\kappa < 0$}
\label{sec:a1}

Here, we show that metric function $f(r)$ vanishes at the singularity $ r_0 := ( -\kappa Q^2/\lambda )^{1/4}$, where $\lambda r_0^4+ \kappa Q^2 = 0$, for $\kappa<0$ and $\lambda > 0$. Let us begin with rewriting $f(r)$ in Eq. \eqref{eq:fr1} as
\be
	f(r)
	=
	\frac{\sqrt{h(r)}}{U(r)}
	\left[
		\int \frac{W(r)}{ \sqrt{h(r)} } dr + 2 \sqrt{ \lambda } M
	\right],
\label{ff_1}
\ee
where
\be
	h(r)
	=
	\lambda r^4 + \kappa Q^2,
\;\;\;
	U(r)
	=
	-\frac{ \lambda r^4 - \kappa Q^2 }{ r },
\;\;\;
	W(r)
	=
	\frac{ ( \Lambda r^4 -r^2+Q^2 )( \lambda r^4 - \kappa Q^2 ) }{ r^4 }.
\ee

Since $h(r)$ vanishes at $ r = r_0 $, the behavior of $f(r)$ near $ r_0$ is not so trivial from Eq. \eqref{ff_1}. However, one can easily know the behavior of $f(r)$ near the singularity if one extracts the pole of the integrand as follows:
\be
	f(r)
	&=&
	\frac{\sqrt{h(r)}}{U(r)}
	\left[
		\int \left( \frac{W(r)}{ \sqrt{h(r)} } -   \frac{W(r_0)}{ \sqrt{h'(r_0) (r-r_0)} } \right) dr
		+ \int \frac{W(r_0)}{ \sqrt{h'(r_0) (r-r_0)} } dr  
		+ 2 \sqrt{ \lambda } M
	\right]
\nn
\\
	&=&
	\frac{\sqrt{h(r)}}{U(r)}
	\left[
		\int Z(r) dr + \frac{2W(r_0)}{ \sqrt{h'(r_0) } } (r-r_0)^{1/2}   + 2 \sqrt{ \lambda } M
	\right],
\label{ff_2}
\ee 
where
\be
	Z(r) = \frac{W(r)}{ \sqrt{h(r)} } -   \frac{W(r_0)}{ \sqrt{h'(r_0) (r-r_0)} }.
\ee
Since we have extracted the pole of the integrand, $Z(r)$ is completely regular at the singularity, which can be expressed as
\be
	Z(r) = - \frac{ W(r_0) h''(r_0) }{ 4h'(r_0)^{3/2} }(r-r_0)^{1/2} + O( (r-r_0)^{3/2} ).
\label{Z_asym}
\ee
Thus, from Eq. \eqref{ff_2} with the help of Eq. \eqref{Z_asym}, we have
\be
	f(r)
	=
	\frac{ \sqrt{ h'(r_0)(r-r_0) } }{ U(r_0) }
	\left[
		2 \sqrt{\lambda} M +  \frac{2W(r_0)}{ \sqrt{h'(r_0) } } (r-r_0)^{1/2}
		+ O((r-r_0))
	\right],
\ee 
which tells us that $f(r) = O( ( r-r_0 )^{1/2} )$ in the vicinity of $r_0$, i.e., $f(r_0)=0$.

\section{Null geodesic}
\label{sec:a2}

Since the Lagrangian is conserved along the geodesic, i.e., $ d{\cal L}/d\lambda = 0 $, one can put ${\cal L} = - \epsilon $, where $\epsilon = 1$ for a massive particle and $\epsilon = 0$ for a massless particle. Thus, we obtain a potential for the radial motion of particles as
\be
	\psi^2\left( \frac{dr}{d\lambda} \right)^2 + V^2 = e^2,
\;\;\;
	V(r) = \sqrt{f\psi^2\left( \epsilon +\frac{\ell^2}{r^2} \right)}.
\ee

Now, let us consider the geodesic of a massless particle ($\epsilon=0$). In this case, rescaling the affine parameter as $\lambda \to \lambda/e$ and introducing the impact parameter $b=\ell / e$, the equation for the radial motion becomes
\be
	\psi^2\left( \frac{dr}{d\lambda} \right)^2 + V^2 = 1,
\;\;\;
	V(r) = \sqrt{ \frac{ b^2 f\psi^2 }{r^2} }.
\ee

The unstable circular orbit (UCO) or the photosphere, of which radius is denoted by $R_{\rm UCO}$, is obtained by solving $V'( R_{\rm UCO} ) = 0$ (and confirming $V''( R_{\rm UCO} ) < 0$, precisely speaking). Then, depending on the value of $V^2(R_{\rm UCO})$, the orbits of a particle are classified into three classes: if $V^2(R_{\rm UCO}) > 1$ the particle infalling from the infinity is scattered off by the potential barrier at some radius larger than $R_{\rm UCO}$; if $V^2(R_{\rm UCO}) < 1$ the particle plunges into the black hole; if $V^2(R_{\rm UCO}) = 1$ the particle circles the black hole infinite times at $r=R_{\rm UCO}$. The condition $V^2(R_{\rm UCO}) \gtreqless 1 $ is equivalent to $ b^2 \gtreqless b_c^2 $, where $b_c$ is a critical impact parameter depending merely on the black hole parameters. We remark that, for the  Reissner-Nordstrom case ($\kappa=0$), $R_{\rm UCO}$ and $b_c$ can be obtained analytically as
\be
	R_{\rm UCO} = \frac{3M}{2} \left( 1+\sqrt{ 1- \frac{8Q^2}{9M^2} } \right),
\;\;\;
	b_c^2
	=
	\frac{ R_{\rm UCO}^4 }{ ( R_{\rm UCO} -r_+ ) ( R_{\rm UCO} -r_- ) },
\ee
where $ r_\pm = M \pm \sqrt{M^2-Q^2} $.


\end{document}